
\input aa.cmm
\newcount\eqnumA\eqnumA=0%
\def\autnumA{\global\advance\eqnumA by 1 {\rm (A \the \eqnumA)}}

\MAINTITLE={The emission spectra of radioweak quasars}
\SUBTITLE={I.  The farinfrared emission}

\AUTHOR={\ts Martina Niemeyer and Peter L. Biermann}

\SENDOFF={\ts M. Niemeyer}

\INSTITUTE={Max-Planck-Institut f\"ur Radioastronomie \newline
Auf dem H\"ugel 69 \newline D-5300 Bonn 1 \newline Germany}

\RECDATE={  }

\ACCDATE={  }

\SUMMARY={We model farinfrared (FIR) spectra of radioweak quasars
with the assumption that the emission is from heated dust, and that
the heating is due to the central engine via energetic particles.
These energetic particles are diffusing from a postulated source near
the central engine through a tenuous galactic halo to arrive at the
dust which is taken to be in molecular clouds in a disk configuration.
This picture does not depend on a particular geometry of the disk
such as warps.  This concept can readily reproduce the range of
observed mm/submm/FIR/IR spectra.}

\KEYWORDS={Interstellar medium:  Cosmic Rays, Clouds, Dust
-- Galaxies: interstellar matter -- Quasars: general -- Infrared:
general}

\THESAURUS={09.03.2, 09.03.1, 09.04.1, 11.09.4, 11.17.3, 13.09.2}

\maketitle

\titlea {Introduction}

The FIR emission from active galactic nuclei was believed for a long
time to be due to nonthermal processes (e.g. Edelson 1986).
The observations of a few nearby Seyfert galaxies (Hildebrandt et al.
1977, Telesco \& Harper 1980, Rieke \& Low 1979, Roche et al.  1984,
Neugebauer et al. 1979, Miley et al. 1984) gave the first hint, that
dust emission might be important as an additional source.  Then, the
very sensitive mm-observations at the 30m IRAM telescope (Chini et al.
1989a,b) combined with the IRAS quasar survey (Neugebauer et al.
1986) turned the trend; the farinfrared emission from radioweak quasars is
now well accepted to be generally due to warm dust (Sanders et al.
1989a, Lawrence et al. 1991, Barvainis 1990).
\par
There are theoretical possibilities to reproduce many, but not all of the
observed spectra with Synchrotron emission from two populations of
relativistic electrons via self-absorption and emission (de Kool et al. 1989,
Schlickeiser et  al. 1991).  However, the observations of very large amounts of
molecular  gas in radioweak quasars (Barvainis et al.1989, Sanders et al. 1988,
Sanders et al. 1989b, Alloin et al. 1992)
demonstrated that the radioweak quasars contain sufficient
gas and dust mass to produce all the emission observed in the FIR.
Hence, our understanding of radioweak quasars now includes an extended region
of
dust clouds, presumably in a disk configuration.  It is less clear what the
energy source of this dust is.  The difficulty lies in the fact that
the FIR emission of radioweak quasars is the strongest of any of the observed
wavelength bands usually, and therefore any primary source must be even
stronger, allowing for some waste and inefficieny.  Another difficulty is, that
we can infer from Plancks law that the radial scale from which the farinfrared
emission is arising, ranges over a fairly large radial scale out to distances
of order $100$ pc, and thus the transport of the energy out to such distances
has to be understood.

First of all, it is tempting to think that the slow
accretion towards the  central object can provide sufficient energy via
dissipation analoguous to  a classical accretion disk; however, it is easily
seen that at the radial scales necessary to emit the power observed of up to
several hundred parsec, the central potential well is much too shallow to
provide so much power.

The second hypothesis, which is actually supported
by the discovery of large amounts of molecular gas, is that a circumnuclear
starburst also provides the FIR emission seen.  It appears as a very reasonable
hypothesis that stars are being formed in these molecular clouds surrounding
the
central engine; a number of nearby Seyfert galaxies and galaxies with extreme
FIR
luminosities clearly demonstrate the activity of such starbursts (a now well
established example is the Seyfert galaxy NGC1068, Wilson et al. 1992).
However,
starbursts have a very clear correlation between their FIR and radio emission,
and also have typical radio spectra:  Normal galaxies have nonthermal radio
emission spectra with flux density $S_{\nu}\,\sim\,\nu^{-0.8 \pm 0.1}$,
sometimes
modified by some thermal free-free emission, or by free-free absorption in a
clumpy interstellar medium, as in the famous starburst galaxy M82 (Kronberg et
al. 1981, 1985).  In M82 the resulting overall radio spectrum is normal at high
frequencies, but curves down to a nearly flat spectrum at low radio
frequencies.  Many radioweak quasars, on the other hand, have a  measureable
radio emission, which is inconsistent with the notion that it arises from the
effects of a starburst; the spectra are rather often flat powerlaws instead of
steep, and sometimes there is evidence that the emission is actually variable
with time.  Furthermore, the ratio of FIR to radio emission is much larger in
many cases for radioweak quasars than for starburst galaxies, clearly requiring
an additional energy source (Chini et al. 1989a).
Resolution of the question, whether extremely intense regions of star formation

also suppress the radio emission relative to the FIR emission, may come from
infrared line observations.
Finally, in some cases the
spatial structure of the radioemission has been determined,  and is found to be
analoguous to radio jets of low power (e.g., Lacy et al. 1992), again
inconsistent with starbursts as the source of the dominant radio emission.
\par
One model which has been advanced (Sanders et al. 1989a) to circumvent all
these problems is the hypothesis that a central X-ray source provides the
power;
this hypothesis neatly ties in with the common interpretation of the X-ray
spectra which strongly suggest reprocessing of primary hard X-rays. These
primary hard X-rays are postulated to originate in a putative central source
above the disk; the energy  source of these X-rays is thought by some to be a
fairly compact region in a putative jet, which accelerates electrons and
positrons in a pair cascade possibly driven by hadronic interaction of
energetic
protons and heavier nuclei with the local photon field.  Such a picture clearly
requires a very large power in this compact region in the putative jet. Given
such a geometry, the X-ray spectra can be modelled in detail (George et al.
1991). In one case, the X-ray emission observed is correlated in its time
variability with
the UV emission (Clavel et al. 1992), thought to arise from the inner region of
an accretion disk, and hence this particular concept requires a power in the
putative jet, driving the energetics of the compact region, of at least the
same
level as the accretion luminosity. The same X-ray source could provide the
power
for the molecular clouds; however, here a difficulty arises from the geometry
of
the situation: The X-ray source cannot easily shine upon the disk evenly from a
few times $10^{16} \;{\rm cm}$ all the way to a few times $10^{20}\;{\rm cm}$.
As a solution warped disks have been proposed, combined with a fair amount of
scattering in the halo region above the disk (Sanders et al. 1989b).  The
warping
has to be rather extreme in order to provide the even illumination at these
large radii.
\par
Here we introduce a different hypothesis that circumvents the
geometric difficulty, but also uses a putative jet and its particle
acceleration:

We propose that the dusty clouds are heated by
relativistic particles. These particles are thought to originate from a source
on the axis of rotational symmetry, possibly a jet or knots in a jet and its
interaction regions with the environment.
Relativistic particles could be accelerated through first order
Fermi acceleration to energies ranging from $10^6$ GeV to $10^9$ GeV by
shocks in the plasma of the jets (Biermann \& Strittmatter 1987, Mannheim
1992).
The difference between radioloud and
radioweak quasars in our picture is not that the jets in radioweak quasars have
a low kinetic luminosity (Lacy et al. 1992), but that their radio emission is
very much weaker, despite their large power.  This could be analoguous to
the weak
nonthermal radio emission of Wolf Rayet stars as compared to the large
nonthermal radio emission of radiosupernovae, which differ probably only in
the velocity of the shockwave which causes the particle injection and
acceleration (see Biermann \& Cassinelli 1993).
This comparison suggests that there is a critical Alfv\'enic Machnumber
for a schock, below which there is very little electron acceleration.
If the analogy to Wolf Rayet stars/ radioradiosupernovae holds, then the
particle acceleration in the jets of radioweak quasars is caused by shocks
with a shock velocity lower than in radioloud quasars.
Thus in our speculative picture
radioweak and radioloud quasars both have jets of strong power,
but in one case the jet power is dissipated fairly deep inside the
host galaxy, and in the other case the dissipation is very far outside
the host galaxy. The source is assumed to be
located above the disk and to transform a fair fraction of the entire source
accretion power into relativistic particles, analoguous to the argument
from Falcke et al. (1993) for the Galactic center source Sgr A$^{\star}$.
They deduce limits
for the important ratio of jet power to disk accretion luminosity
for the Galactic center source and found that the data lead to a lower limit
for
this ratio of order $0.1$. Rawlings \& Saunders (1991) came to a similar
conclusion,
that the extracted energy by the jet is comparable to the energy radiated by
the
accretion disk, from a purely observational analysis of jet and disk emission
of a large variety of radio galaxies.
Such a concept is consistent with
the model for strong jets discussed e.g. by Camenzind (1990).
Thus, in our model we assume that the compact jet causes particle acceleration
at a level of power comparable to the photon luminosity.
In such a picture the occasionally observed UV -- X-ray correlation is caused
by the particle heating of the disk, where the particles originate
in the jet, which in turn connects to the inner edge of the accretion disk.
The particles
easily scatter in the magnetic plasma that permeates the halo above the disk
all
the way from the central engine.  And since dust clouds of the column densities
typical for central regions of galaxies readily provide sinks for energetic
particles, they can be heated.
\par
There are a number of important
questions to be answered: First, we have to demonstrate that this
mechanism can produce the observed FIR spectra with reasonable assumptions
about
the scattering of energetic particles in the halo region.  Second, we have to
discuss how the particles heat not only the outer part of a disk where the
interstellar medium consists of clouds and an intercloud medium, but also in
the
innermost part of the disk which is dense hot gas close to a central compact
object like a rotating black hole; this leads to a detailed treatment of the
X-ray emission. Third, we have to deal in detail with the actual transfer of
energy from the particles to the interstellar gas; there are several ways of
doing this, a) p-p collisions are clearly an important ingredient; but, b) the
excitation of Alfv\'en waves which are damped in the mixed neutral-ion gas
of the dense dust clouds is also an important process (Skilling  \& Strong
1976,
Kulsrud \& Pearce 1969)
which may have also relevance for the inner part of our Galaxy.  Fourth, we
have
to check the overall model by calculating the Gamma- and Neutrino background
due
to all cosmological radioweak quasars.  The present experimental limits on both
backgrounds are already rather restrictive for any such model.  A very brief
outline of the physical arguments contained in our quasar model with emphasis
on
the neutrino emission was presented at the Hawaii meeting (Biermann 1992), and
consequences of the effect of strong neutrino emission on stars and stellar
evolution were published in MacDonald et al. (1991), while the dust spectra
were
first presented by Niemeyer in her M.Sc. thesis (1991).
The resulting neutrino spectra and the cosmological neutrino background due
to radioweak quasars have been calculated by Nellen et al. (1993).
\par
In this paper we propose to take the first step and test our basic
hypothesis by calculating the FIR spectra that can be produced in this
model, and compare them to observations.  The basic model parameter is
the radial behaviour of the diffusion coefficient that governs the
transport of the energetic particles from the putative central source
out to the dust clouds in the disk.
\par
The paper is structured as follows.  In section 2 we discuss the
physical ingredients to our model, in section 3 we describe the model in
detail, in section 4 we compare the resulting FIR spectra with
observations, and in the final section 5 we draw conclusions from our
model fits and speculate on possible generalizations of our model.

\titlea {The basic physical ingredients}

Our basic concept devolves from the assumption that all active
galactic nuclei have an inner accretion disk with a jet that starts
near the inner edge of the disk and produces strong shocks inside
itself and near its boundaries.  This jet is assumed to carry a kinetic
luminosity similar to the accretion luminosity. As noted above, this
assumption finds support in the recent interpretation of the
Galactic center radio source Sgr $A^{\star}$. Energetic particles gain
energy in these shocked regions (e.g. by first order Fermi acceleration, see
Drury 1983) and diffuse out into a region of tenuous gas which permeates the
entire zone above the disk of the host galaxy, from the jet on out.  Support
for such a picture is given by the observation that even radioweak quasars
often have radio spectra that are flat ($S_{\nu}\sim \nu^{\alpha}, \alpha
\simeq 0$) sometimes variable, although very weak in their emission level (a
summary of such observations is given in Chini et al. 1989a).  Such spectra
are analoguous to the radio spectra of radio quasars, and can be
interpreted in a similar way (Lacy et al. 1992, Falcke et al. 1992).
We assume in our picture that there is a
population of accelerated nuclei - as opposed to electrons or
positrons - which contains an appreciable proportion of the total
accretion power.  Thus, our picture drains most of the putative jet power
fairly deep inside the mother galaxy, similar to Fanaroff Riley class I radio
galaxies.
\par
Given then a strong source of energetic particles somewhere near
the central engine, these particles diffuse out through the tenuous plasma
which permeates the region above (and below) the inner and outer disk;
the parameters of this diffusion are the main uncertainties of our
model and will have to be fitted from the observations.  The basic
test is then whether any reasonable model for this diffusion gives a
physical explanation for the observed spectra.
We note that with a diffusive distribution of the energy we can get nearly all
of the power of energetic particles to heat the disk
in contrast to a radiation model which reaches only a small covering fraction.
\par
The details of the model then break up into three parts, first, the
diffusion of energetic particles through the halo and the energy
deposition in the dusty disk, second, a detailed treatment of
radiative transfer of the infrared radiation out of the disk, and
third, a presentation of the numerical results.

\titlea {The detailed model}

\titleb {The diffusion through the halo and energy deposition in the disk}

The derivation of the general diffusion equation applicable here has
been given by Chandrasekhar (1943).

$${\vec \nabla} (D({\vec r},E){\vec \nabla} W({\vec r},E)) \, + \,
Q({\vec r},E) \;= \;0.  \eqno\autnum $$
The source can be either a
point source or a line source in the central axis of symmetry.  We
neglect here further energy gain or loss of the particles as well as
overall convection.  These assumptions result in restrictions of our
model which we discuss below.  Furthermore, $W({\vec r},E)$ is the
distribution function which we have to determine, $D({\vec r},E)$ the
assumed diffusion coefficient and $Q({\vec r},E)$ is the source
function.  The boundary conditions for the solution of the diffusion
equation are first, the conditions at infinity, and second the
conditions on the disk:

At infinity the distribution function has to converge to zero

$$W|_{\infty} \;=\; 0 , \eqno\autnum $$
while on the disk we assume
total absorption

$$W|_{\rm disk} \;=\; 0. \eqno\autnum $$
This assumption does not
mean that an energetic particle has to be absorbed each and every time
it first hits the disk, but rather that its mean free path near the
disk in the halo is so short, that it scatters through the disk often
enough to be absorbed before it leaves the region near the disk again
on its repeated scatterings; however, the particles should not just get
absorbed in the skin of the dusty disk, and so their mean free path should not
be much smaller than the disk thickness.  Since the disk is made up of a
large assembly of dense molecular clouds, this is plausible.
\par
Given a solution to our diffusion equation we derive the
energy deposition rate as a function of particle energy (Chandrasekhar 1943)
from

$$q({\vec r},E) \;=\; -D({\vec r},E)({\vec \nabla}\,W({\vec
r},E))|_{W=0}. \eqno\autnum $$
With that solution the total energy
deposition is given by the integral over all relevant particle
energies

$$F_{\rm in} \;=\; \int_{E_{\rm min}}^{E_{\rm max}} q(\vec
r,E)\,E\;dE. \eqno\autnum $$
The magnetic field structure may have a variety of
behaviours, e.g. a solar wind type Parker spiral, where the radial behaviour of
the magnetic field is either with an exponent of $-1$, in the outer
regime of the Archimedian spiral and not very near the pole, or with an
exponent
of $-2$, in the inner part of the spiral, where the magnetic field
is still corotating, or near the pole regions, where the magnetic field is
radial.
\par
The diffusion coefficient can be derived from
various physical processes:

First, resonant scattering of relativistic
particles in Alfv\'en waves, which leads to a diffusion coefficient of

$$D \;=\;{1 \over 3}\,r_g\, c\,{{B^2/{8 \pi}} \over {I(k)k}}
,\eqno\autnum $$
where the Larmor radius $r_g$ is for relativistic
particles of energy $E$ given by

$$r_g \;=\;{ E \over {e c B}} ,\eqno\autnum $$
and $I(k)$ is the
energy density of resonant Alfv\'en waves per wavenumber $k$.  In the limit of
strong turbulence the waves have an energy
density of similar intensity as the underlying magnetic field, and the
diffusion
coefficient converges to the socalled Bohm limit of

$$D \;=\;{1 \over 3}\,r_g\, c.\eqno\autnum $$
Parallel to the magnetic field this Bohm diffusion coefficient is the lower
limit to the actual diffusion coefficient, and perpendicular to the magnetic
field it is an upper limit.  Thus, given an underlying Parker spiral for
instance the diffusion from the halo to the disk is strongly inhibited,
since it requires diffusion perpendicular to the magnetic field lines. If one
were to assume that the magnetic field is highly turbulent, so that such a
diffusion is a proper description in all three spatial dimensions over the
length scales of interest, then the Bohm limit is a possible approximation to
the actual diffusion coefficient, and diffusion towards the disk is still
strongly limited. Hence, Bohm like diffusion is unlikely to play a role in the
energetic particle transport, as is required by our picture of the inner
regions of quasars.
\par
Second, we might have fast convective motions, which
transport embedded energetic particles diffusively.  Convective
turbulence is independent of particle energy, provided its internal length
scale is larger than any relevant Larmor radius, and scales with the local
length scale of the system and the velocity of the convection.  Such turbulence
may well describe better diffusion perpendicular to the underlying magnetic
field.  If the convective diffusion coefficient is selfsimilar with radius $r$
in
the sense, that its radial scale is linear with $r$ and its velocity scale is
constant with $r$, then a radial dependence of the diffusion coefficient with
exponent $1$ is inferred.
The analogy with the diffusive fast convection connected with shocks in stellar
winds or in the homogeneous interstellar medium is useful:  Supernova shocks
demonstrate from radio observations that the magnetic field structure is mostly
radial all the way around the observed perimeter (Dickel et al. 1991), and this
behaviour is most readily interpreted as the result of fast convective
turbulence in the shock region, with a length scale proportional to the radius
itself (Biermann \& Strom 1993). The smallest dominant length scale is the
thickness of the shocked layer:  In stellar winds this length scale is $r/4$
and
for shocks in a homogeneous medium this length scale is $r/12$ (Biermann 1993,
Biermann \& Cassinelli 1993, Biermann \& Strom 1993). For substructure in the
shocked layers the scale can be smaller as shown by observations of supernova
remnants (Biermann \& Strom 1993), of order $0.01\,r$.  These analogies imply
an upper limit to the mean free path of order $0.1$ the radial distance $r$,
and
probably larger than $10^{-3} \,r$.  Hence each energetic particle that arrives
near the disk at all, is ultimately absorbed near its location of first
arrival.
In such a
picture the diffusion coefficient is independent of particle energy and would
scale both with the inherent length scale and velocity scale.

We can estimate a numerical value for the diffusion coefficient as follows.
The
diffusion time scale has to be shorter than the convective time scale of any
overall halo wind of velocity $V_W$.  This entails the condition

$${r^2 \over D} \;<\;{r \over V_W}. \eqno\autnum $$
This can be transformed with writing the mean free path as $\epsilon \,r$
and

$$D \;=\;{1 \over 3} \,\epsilon \,r \,v_t ,\eqno\autnum $$
where the $v_t$ is the turbulent convective velocity to the condition

$$v_t \;>\;{3 \over \epsilon}\,V_W.\eqno\autnum $$
Since the turbulent velocities in the gas can be estimated from narrow emission
line width to be of order $10^3$ km/sec, we thus obtain a numerical estimate
for
the diffusion coefficient of

$$D \;=\;10^{24\pm 1 }\,{r \over {\rm pc}}\,{\rm cm^2\,s^{-1}} ,\eqno\autnum $$
using our estimate above for the mean free path.  Here any further
dependence on radius $r$ is hidden in the uncertainty of the numerical
factor.  This diffusion coefficient is independent of energy; it may have to be
increased in the inner region, where the X-ray emission is produced, by a
factor
of order $10$.  The inequality for the wind velocity implies that any overall
wind at the relevant radial scales is much less than the turbulent velocity.

Independent of any such particular model for the diffusion
coefficient, we assume here for the diffusion coefficient a general powerlaw
dependence on spherical radius $r$ with either an exponent $
\beta \,=\,0,\,1$
or
$2$:

$$D(\vec r,E) \;=\; D_1\,r^\beta , \eqno\autnum $$
and infer then the
powerlaw exponent from a fit to observation.

In the following we do not consider losses, e.g. radiation losses and
convection, which would change the energy distribution of the particles.

It will be shown below that the
spectrum itself is independent of the numerical value of the diffusion
coefficient, since we have assumed a stationary state.  However, for these
basic assumptions to be justified, that lead to a stationary state, the
diffusion coefficient has to be quite large and is numerically estimated
above.
\par
Next, we describe the method of solution of the partial differential
diffusion equation in a stationary state.

\titlec {Point source }

Since we assume a dependence of the diffusion coefficient on the
radial distance $r$, we have to solve the partial differential
diffusion equation in spherical coordinates.

$$\eqalignno{-Q({\bf r},E)/D_1 &= \beta r^{\beta-1}\cr &+
r^{\beta-2}\left({\partial \over
\partial r}\left( r^2{\partial W\over \partial r}\right)+{\partial
\over \partial \mu}(1-\mu^2){\partial W\over \partial \mu}\right)
&\autnum \cr}$$
with

$$Q({\bf r},E) = {1\over 2\pi z_0^2}\delta(\mu-1)
\delta(r-z_0)\,Q_2(E) \eqno\autnum$$
where $Q_2(E)$ is the energy
distribution of the energetic particles after leaving the source. We
use a powerlaw here since diffuse shock acceleration usually gives
powerlaw spectra (e.g., Drury 1983) for the energetic nuclei in the
intervall of interest 1 to order $10^8$ GeV (see Biermann \& Strittmatter
1987, Mannheim 1992) :

$$Q_2(E) = A_p E^{-\gamma}, \;\;\;\; {\rm with}\; \gamma=2. \eqno\autnum$$
The constant of normalization $A_p$ can be expressed with the total power
that goes into the energetic particle channel $L_p$:

$$\eqalignno{L_p &= 4\pi\,\int_{E_{\rm min}}^{E_{\rm
max}}\int_0^{2\pi}\int_{-1}^1
\int_0^\infty Q({\bf r},E) E\,dE\,d\mu\,d\phi\,r^2\,dr\cr
& \cr &= A_p\,4\pi\,\ln(E_{\rm max}/E_{\rm min}) & \autnum \cr}$$
\par
As demonstrated in Appendix A, where we use the scattering time method
(e.g. Wang and Schlickeiser 1987) to solve the partial differential
equation, the distribution function has the general form:

$$W(r,\mu ) = -\sum_{n=0}^\infty {C_{2n+1}P_{2n+1}(\mu) t_{2n+1}(r)}
\eqno\autnum$$
with

$$t_{2n+1}(r) =-{1\over 2\pi d D_1}\,Q_2(E) \cases{
z_0^{d_2}\;r^{d_1}, &if $0\le r\le z_0$\cr \cr z_0^{d_1}\;r^{d_2}, &if
$z_0\le r\le \infty.$\cr}
\eqno\autnum$$

$P_{2n+1}(\mu)$ are the Legendre polynomials and the constant
$C_{2n+1}$ follows from the boundary conditions, $n \in {\bf N}$.
$d,d_1,d_2$ are given in Eq.((A24),(A25),(A26)).

The deposition rate of the particles is determined by Eq.(4).
The partial derivative with respect to $\mu$ gives for $q(r)$:

$$\eqalignno{q(r,E) &= \sum_{n=0}^\infty {C_{2n+1}\over 2\pi
\,d}\,{\partial P_{2n+1}(\mu )\over \partial \mu}|_{\mu=0}\cr & \cr \cr
&*Q_2(E)\cases{ z_0^{d_2}\;r^{-1+\beta +d_1} &if $0\le r \le z_0$\cr \cr
z_0^{d_1}\,r^{-1+\beta +d_2}&if $z_0\le r \le \infty$\cr} & \autnum
\cr}$$
This is then the full solution for the point source.  For the
deposited energy flux we have after integrating over the particle
energy:

$$F_{\rm in}(r) = \sum_{n=0}^\infty B_{2n+1}\,L_p
\cases{ z_0^{d_2}\;r^{-1+\beta +d_1} &if $0\le r \le z_0$\cr \cr
z_0^{d_1}\,r^{-1+\beta +d_2}&if $z_0\le r \le \infty$\cr}
\eqno\autnum $$

with

$$B_{2n+1}={C_{2n+1}\over 2\pi\,d} {\partial P_{2n+1}(\mu)\over
\partial \mu}|_{\mu=0}.\eqno\autnum$$

\titlec {The line source}

Here we consider the possibility that the source is extended along the
symmetry axis z; this case can be considered as a limiting addition of
many point sources.  The physical concept is given by many knots in a jet.  For
the distribution function we then have:

$$ G(r,\mu ) =\int_{z_{\rm min}}^{z_{\rm max}}
W(r,\mu)\,\left({z_0\over z_{\rm min}}\right)^{-\alpha_1}\,dz_0,
\eqno\autnum$$

$z_{\rm min}$ and $z_{\rm max}$ denote the limits of the extent of the
source.
\smallskip

\noindent The deposition
rate of particles is then analoguously to the case considered earlier:

$$q_z(r, E) = -D(r)\left( \nabla
G(r,\mu)\right)|_{\mu=0}\eqno\autnum$$

\noindent The deposited energy flux then is obtaining after integrating over
$E$:

$$F_{\rm in}(r) = \int_{E_{\rm min}}^{E_{\rm max}}
q_z(r,E)\,E\;dE
\eqno\autnum$$

Two examples of results for the energy deposition rates are
shown in Fig.(1) and (2).
\smallskip

The first figure shows the energy deposition of a line source with
an intensity $z^{-1}$ and three
different diffusion coefficients. The second figure shows the energy deposition
rates for the three different types of sources with a diffusion coefficient
$D = D_1r$.\par

\begfig{5.5}cm{\figure {1}{Comparison of the energy deposition rate
in the case of a line source, with an intensity $z^{-1}$ and three
 different diffusion coefficients.
 ($L_p=10^{45}\;$ erg/s, $z_{\rm min}=0.1$ pc).}}
\endfig

\begfig{5.5}cm{\figure {2}{Energy deposition rate, with $D =D_1\,r$.
{\bf q1} -- point source; ($L_p = 10^{45}{\rm erg/s},\,z_0=0.1$pc);
{\bf q2} -- line source, with an intensity $z^{-1}$,
{\bf q3} -- line source, with an intensity $z^{-1.5}$;
($L_z=10^{45}{\rm erg/s},\,
z_{\rm min}=0.1$pc).}}
\endfig

It can be pointed out that the $r$-dependence of the energy deposition
can be approximated by
powerlaws over a large range of radius $r$. The steepness of the curves depend
on two parameters:
\item{a)} The greater the radial dependence of the diffusion coefficient,
the flatter the curves.
\item{b)} The greater $\alpha_1$, the steeper the curves.\par
\smallskip

\titleb {Radiative transfer in the dusty disk}

Dust particles are of mixed chemical composition with different grain
size distributions.  Mathis et al. (1977, hereafter referred to as MNR)
developed a dust model for the composition, absorption and emission of
interstellar dust.  They succeeded to approximate the dust grain size
distribution with a powerlaw over the range in wavelengths
$0.11<\lambda [{\rm \mu m}]<1$:

$$ f_i(a)=K_i\, a^{-3.5}\;\;{\rm for}\;\;0.005{\rm \mu m}\le a\le 1{\rm \mu m}
\eqno\autnum$$
where $f_i(a)$ is the number of dust particles of one kind with size
between $a$ and $a+da$, and $K_i$ is a constant.
This powerlaw can be interpreted as a quasistationary fragmentation
(Biermann and Harwit 1980).
There are some recent dust models which we discuss in the next
subsection (4.1.6).\par

With this dust grain size distribution the optical depth
to absorption by dust is:

$$\tau_{\lambda,s}=\sum_i \int \, ds \int_{a_{\rm min}}^{a_{\rm max}}
f_i(a)\sigma_{\lambda,i}(a)\;da \eqno\autnum$$
with $a_{\rm min},\;
a_{\rm max}$ the minimum and maximum dust grain size.
$\sigma_{\lambda,i}(a)$ is the absorption cross section dependent from
the kind and grain size of dust.\par
\smallskip

\noindent For the calculation of the FIR spectrum, one must determine
the dust temperature. Here we discuss the case when the dust
temperature is determined by radiative processe alone. So we must
construct a heat balance that takes into account absorption and
emission of energy.\par

\titlec {Absorbed Energy}

In this paper we do not discuss in detail how the energy is
transformed from an impinging flux of energetic particles into an
effective gas and dust heating.  There are two channels which are
clearly important.\par
\smallskip

a) The p - p or nucleus - nucleus collisions which
transform particle energy into pions (and neutrinos, irrelevant in the
context of this paper, but discussed in detail in Nellen et al. 1992), which in
turn decay and finally deposit their energy in electrons and positrons as well
as
gamma-photons of a range of energies.  These electrons/positrons and gammas in
turn thermalize their energy by further encounters with atoms and their shells
(see, e.g., Spitzer \& Tomasko 1968).

b)  The second process is the
excitation of Alfv\'en waves by a gradient of the particle
distribution, which is necessarily formed when many particles are
absorbed in p - p and corresponding collisions.  This wave excitation
in turn can lead to a very strong wavefield, which dissipates readily
in a neutral-ion plasma inside dense dust clouds; it would
dissipate considerably less in a fully ionized medium, such as the
innermost accretion disk.  These processes we will discuss
elsewhere.\par
\smallskip

Here we simply assume that a fixed fraction of the energy of the
impinging particles is deposited as heat. This leads to:

$$W_a^D = 4\,\delta_F\,F_{\rm in}(r)
\eqno\autnum$$
where $F_{\rm in}(r)$ is the total energy depositon in the disk and
$\delta_F<1$ the scale factor.  We assume in the following that this
efficiency is $1/3$, since this is the proportion of impinging
particle energy that goes into electrons or positrons rather than
energetic photons or neutrinos.  Energetic photons and neutrinos are
not as easily thermalized as electrons or positrons and are therefore
ignored here.
\par

\titlec {Emitted Energy}

The emitted energy is:
$$W_e^D = 4\pi\,
\int_0^{\infty}B_{\lambda}(T_D)\,(1-\exp(-\tau_{\lambda}))\;d\lambda
\eqno\autnum$$
where $\tau_{\lambda}$ is the optical depth and $B_{\lambda}(T_D)$
the Planck function.

Knowing the optical depth for dust one can compute the
integral (Eq.(29)). \par
Here we introduce the assumption that the dust
properties in such central region are basically the same as in the solar
neighborhood.  This means we will use the cross section of dust
particles integrated over the grain size distribution.
Mezger et al. (1982) give the result of such an integration,
the total cross
section of dust to absorption, for silicates and graphites in the form
of a cross section per hydrogen atom at a given wavelength
$\sigma_{\lambda,H}$.  The absorption cross section is derived in a
combination of observational and theoretical arguments and can be
fitted by a combination of powerlaws and parabolas.
The optical depth is then given by:

$$\tau_{\lambda} = \sigma_{\lambda,H}\;N_{H}(r),\eqno\autnum$$
where $N_H(r)$ is the hydrogen column density.
We ignore the substructure in thermal properties of the gas and dust
perpendicular
to the disk.\par

One important function which we have to assume is the radial
variation of the column density. We use:

$$N_{H}(r) = N_{H,0}\,\exp(-r/r_c),\eqno\autnum$$
where $r_c$ is a characteristic scale.
$\sigma_{\lambda,H}$ is the
absorption cross section per hydrogen atom as a function
of the wavelength. Using the experimentally tested opacity
approximation of Mezger et al. (1982) mean that we actually
independent of the particular dust grain size distribution.
The optical depth decreases outwards.\par

\titlec{The dust temperature $T_D$}

After considering the emission and absorption of dust, we proceed now
to a determination of the dust temperature by assuming that emission
and absorption balance.

$$W_e^D(T_D) =W_a^D(r). \eqno\autnum$$

\begfig{5.5}cm{\figure{3}{Temperature distribution of the point source with
different diffusion coefficient. ($L_p=10^{45}{\rm
erg/s},\,z_0=0.1$pc).}}
\endfig

Since the incoming flux
of energy depends on radius in the disk geometry considered, the dust
temperature is also dependent on radius $r$. The following figures
show the temperature distribution for several sources which we discuss
in detail.

Fig. (3) shows the temperature distribution of the point source
dependent on the spherical radius $r$ with different diffusion
coefficient.  As already apparent in the earlier expression
for the radial dependence for the energy flux impinging on the disk,
the dust temperature can be closely approximated by powerlaws.
This powerlaw is valid over a large range of the radii considered.
It can be pointed out that the steepness of the curves decreases
with increasing radial dependence of
the diffusion coefficient according to the total energy deposition
rate.
With increasing radial dependence of the diffusion coefficient
more particles reach the outer part of the disk to heat the dust.

\begfig{5.5}cm{\figure{4}{Temperature distribution of the line source
with an intensity $z^{-1}$; variation of the diffusion coefficient.
($L_z=10^{45}\,{\rm erg/s},\; z_{\rm min}=0.1$pc)}}
\endfig

Fig. (4) shows the temperature of a line source with an intensity
$z^{-1}$ with different diffusion coefficients.  The same features
appear, but in this case the curves run together at large radii.
It can be established that the solutions for a line source approximate
the solution for a point source in the limit of steep z--dependence
$\alpha_1$,
on the other side the curves will be flatter if the z--dependence
$\alpha_1$ is smaller, independent on the diffusion coefficient.
\par

\titleb{Calculation of the FIR spectrum}
\titlec{Radiative transfer in the FIR}

We use the essential elements of the radiative transfer calculations.
We assume first thermodynamic equilibrium and second that the
temperature varies only negligibly parallel to the symmetry axis $z$,
the solution of the transport equation is:

$$I_{\nu} = B_{\nu}(T)\,\left(1-\exp(-\tau_{\nu})\right).\eqno\autnum$$
$\tau_{\nu}$ is the optical depth through the disk parallel to the
$z$-axis and $B_\nu(T)$ the Planck function.  In the next step we have
to integrate over the entire disk to obtain the luminosity:

$$L_{\nu} = 4\pi\int_{r_{\rm min}}^{r_{\rm max}}
B_{\nu}(T)\,\left(1-\exp(-\tau_{\nu})\right)\,2\pi\,r\;dr.\eqno\autnum$$\par

\titlec {Numerical results}

Here we calculate the FIR spectrum of a disk in the wavelength range
$10$ und $1300 \,{\rm\mu m}$. The most important characteristica to be
interpreted are the rapid rise from the mm-wavelengths to the FIR
$L_{\nu}\sim \nu^\alpha$ with $\alpha \ge 2.5$, and, on the other
hand, the slow decrease with frequency beyond the farinfared with a
spectral index $\alpha\simeq -1$.

With Eq.(34) we calculate the
luminosity for a finite disk.  The limits of integration are the
minimal and maximal radius in the disk contributing.  Here the maximal
radius is given by the transition from energetic particle heating due
to the central engine to heating through various channels by young and
massive stars.  This happens at large radii, where the optical depth
is small, where we have a dust temperature of ($T_{D,{\rm min}} \simeq
20$K).  The minimal radius can be defined by the sublimation
temperature of dust ($T_{D,{\rm max}} \simeq 1500$K).  In some cases the
maximal temperature is lower, because the sublimation temperature is not
reached.
We use the same assumption as in the subsections above, that
the dust properties in such central regions are basically the same as
in the solar neighborhood, so the optical depth is given by Eq.(30).

The following figures show the results for several sources which we
discuss in detail with the three cases of the diffusion coefficient.
Fig. (5) demonstrates the case of a point source arbitarily taken to
be at $0.1$ pc above/below the disk in the symmetry axis.  All three
cases of the radial behaviour of the diffusion coefficient show the
steep increase from the mm-range towards the FIR, as required by the
observations. The emission through the IR towards the NIR decreases
very slowly. In the case of the constant diffusion coefficient
there is even a slow increase in the IR--range.
The reason for this characteristic is the radial
distribution of the dust temperature $T_D(r)$. The steeper the radial
dependence of the temperature, the greater the spectral indices
in the IR--range. The radial dependence of the temperature is a function of
the radial dependence of the diffusion coefficient and the source
function.

\begfig{5.5}cm{\figure{5}{Spectra for a point source, with variation of
the diffusion coefficient. ($L_p=10^{46}\,{\rm erg/s}$, $N_{H,0}=
10^{24}\,{\rm cm^{-2}}$, $z_0=0.1\,{\rm pc}$)}}
\endfig

\noindent In Fig.(6),(7) we show the cases where we have a line source,
with the source power decreasing as $z^{-\alpha_1}$.  Again we see a
strong increase of the emission from the mm-range towards the FIR, but
in contrast to the case of a point source the decline through the FIR
towards the NIR is stronger; the most extreme case is for the line
source with an intensity $z^{-1}$ and $D=D_1r^2$.
Again, in the limit of steep
z--dependence the solution for a line source approximate the solution
for a point source.

\begfig{5.5}cm{\figure{6}{Spectra for the case of a line source with
intensity $z^{-1}$ and variation of the diffusion coefficient.
($L_z=1.25*10^{46}\,{\rm erg/s}$, $N_{H,0}=10^{24}\,{\rm cm^{-2}}$, $z_{\rm
min} = 0.1\,{\rm pc}$)}} \endfig

\begfig{5.5}cm{\figure{7}{Spectra for the case of a line source with
intensity $z^{-1.5}$ and variation of the diffusion coefficient.
( $L_z=10^{46}\,{\rm erg/s}$, $N_{H,0}=10^{24}\,{\rm cm^{-2}}$, $z_{\rm
min}= 0.1\,{\rm pc}$) }}
\endfig

We can summarize by noting that one kind of the
calculated spectra show a good representation of what is seen in the
observations.  In the mm to FIR wavelength range the local spectral
index $\alpha$ is always $\ge 2.5$, independent of the specific
dependence of the diffusion coefficient on radius.  On the other hand,
the spectra in the IR to NIR range are indeed strongly dependent on
the source function and on the specific law for the radial dependence
of the diffusion coefficient chosen. The decline through the IR gets
stronger with an increasing radial dependence of the diffusion
coefficient.
These Figures ((5),(6),(7)) demonstrate that only the spectra of a line
source with an intensity $z^{-1}$ show the slope as the observation requires.
The observation show in the IR--range power laws with a spectral index
$\alpha\sim 1$.
All other sources with $\alpha_1 \ge 1$ can be ignored, because
the decline in the IR--range of the model spectra are too flat.
There is no clear restriction to one particular
radial behaviour of the diffusion coefficient, because the difference
in the decline of the IR--range is only small.
If there are some observed spectra with an IR--spectral index $\alpha \ge 1$
it is possible to use lines sources with $\alpha_1 \le 1$ for modelling.
The examination of this will be done in the next section.
\par

\titlea {Comparison with observations}
\titleb {The influence of the model parameters}

Here we discuss in the first subsection the influence of varying the
model parameters on the resulting infrared spectra, and in the second
subsection we actually discuss the fits to the observations and the
consequences.

Our model parameters are the following:
\item{a)} the distance between point source or minimum distance of a
line source and the symmetry center $z_0$ or $z_{\rm min}$;
\item{b)} the column density of neutral hydrogen $N_{H}(r)$, which
is
directly proportional to the optical depth $\tau_{\nu}$;
\item{c)} the luminosity of the source $L_Q$ ;
\item{d)} the exponent $\gamma$ in the energy spectrum of the particles;
\item{e)} the radial behaviour of the diffusion coeffcient;
\item{f)} the radial length scale $r_c$ of the neutral hydrogen column
density;
\item{g)} the amount of the total energy transformed into the dust heating;
\item{h)} the dust model

There are hidden parameters in the actual value of the diffusion
coefficient and in the relation between neutral hydrogen column
density and heavy element abundances. It is obvious, that the latter
introduces only a scaling in the optical depth, while the former is
discussed above.

We only use the model spectra of a line source with an intensity $z^{-1}$
for discussing the effect of varying the model parameter, because only this
source  function shows the required slopes in the FIR-- and IR--range.
\par

\titlec{Variation of $z_0$ or $z_{\rm min}$}
\begfig{5.5}cm{\figure{8}{Spectra for the case of a line source
with an intensity $z^{-1}$; variation of the
distance from the center of symmetry. ($L_z = 1.25*10^{46}\,{\rm erg/s}$,
z1 -- $z_{\rm min}=1\,{\rm pc}$, z2 -- $z_{\rm min}=0.1\,{\rm pc}$, z3 --
$z_{\rm min}=0.01\,{\rm pc}$}} \endfig

In this section we discuss the influence of the distance between a
initial point of a line source and the
symmetry center, at the center of the disk.

Fig.(8) shows the
spectra calculated for different $z_{\rm min}$.
The spectra are only shifted in the
$L_{\nu}-\nu$--diagram to lower luminosity and lower frequency, if
$z_{\rm min}$ decreases vice versa.
In the case where $z_{\rm min}= 1$pc appears a bend in the IR--range
of the spectrum, because of the low maximal dust temperature of $400$K,
for this distance the source luminosity is not high enough
to heat the dust to higher temperature.
We conclude that the location of the source on the $z$--axis is
irrelevant, as long as this distance is smaller than a maximal distance.
This distance is dependent on the source luminosity, it increases
with increasing source luminosity.
It is important to
note that there is no minimum distance the source has to have.
This is because with decreasing minimum distance more particles are
deposited in the inner part of the disk, but the radial
dependence of the temperature
is equal in the outer part of the disk. The
deposition of more energetic particles in the inner part of the
disk the maximal IR--luminosity
decrease with decreasing minimal distance $z_{\rm min}$.
The relation can be described by a logarithmic scale factor:

If the minimal distance $z_{\rm min}$ decreases by a factor
$X$, the IR--luminosity decreases by a factor $Y=\ln X$ for a given source
power.

\par

\titlec{Variation of the source luminosity}

By the variation of the source luminosity we get also a shift in the
$L_{\nu}-\nu$--diagram.  Fig.(9) shows the variation of $L_z$ for a
line source with an intensity  $z^{-1}$ and $D\sim r$.
If the source luminosity increases the spectrum shifts to higher
luminosity and higher frequency.

\begfig{5.5}cm{\figure{9}{Spectra for the case of a line source
with an intensity $z^{-1}$; variation of the
source luminosity. (Parameters: $D\,=\,D_1\,r$, $z_{\rm min}=0.1$pc,
$N_{H,0}=10^{24}$cm$^{-2}$, L1=$1.25*10^{45}$ erg/s, L2=$1.25*10^{46}$ erg/s,
L3 =$1.25*10^{47}$ erg/s) }}\endfig

\titlec{Variation of the exponent $\gamma$}

In this paper we do not discuss in detail how the particle energy
actually gets transferred to the molecular gas and dust. For the case
of p--p--interaction and the resulting thermalization of secondaries
it is easy to see that approximately equal amounts of energy get
channeled into neutrinos, gamma photons and electron/positron pairs.

In this paper we always use $\gamma=2$. Changing this value of the
powerlaw energy distribution of the particles can lead to different
proportions of resulting secondary particles due to the threshold
involved in the detailed steps. This is beyond the scope of this paper and
would translate only into a different normalization between source
power and particles distribution normalization.  We ignore this in the
following.

\par

\titlec {Variation of the column density $N_H(r)$}
\titled {a) Variation of the scale of the column density}\par

\noindent The column density is described by Eq.(31).  A variation influence
the
optical depth. This in turn changes the resulting spectra, as shown in
the following Fig.(10).
First the spectra shift in the $L_\nu-\nu$-diagram, especially the
maxima of the spectra shift with decreasing column density to higher
frequencies.  On the other hand, the spectral index of the steep drop
off in the FIR/mm-range remains the same.

\begfig{5.5}cm{\figure{10}{Spectra for the case of a line source
with an intensity $z^{-1}$ with
different scale $N_{H,0}$ of the column density.
(Parameters: $L_z=1.25*10^{46}
\,{\rm erg/s},\;z_{\rm min}=0.1pc,\; D(r)=D_1r^1$, N1=$10^{25}$ cm$^{-2}$,
N2=$10^{24}$ cm$^{-2}$, N3=$10^{23}$ cm$^{-2}$).}}\endfig

Second, we note that for high values of column density $N_{H,0}$ the
spectra are identical in the IR-range, with a drop off in the
FIR-range; the characteristic frequency of this drop off is higher
for smaller column density.  Below a critical value for the column
density the spectra no longer resemble each other very well.  The
spectra are no longer smooth, and can no longer be described by
single powerlaws.  The maximum just beyond the steep drop off in the
FIR/mm-range at higher frequencies is near $6.5*10^{12}$ Hz.  On the low
frequency side of this maximum the spectra do not drop off as steeply at first,
and only approximate the steep drop off seen in the spectra for higher column
densities below $2.5*10^{12}$ Hz. \par

For the case of the lowest column density, the IR spectrum is similar
to the case for the higher values of the column density over only a
small range in frequency. Beyond the maximum near $6.5*10^{12}$ Hz
there is a decline to higer frequency which is flatter. The curves run
together at roughly $3*10^{13}$ Hz.

The reason for this behaviour is the dependence of the optical depth
from the absorption cross section $\sigma_{\nu,H}$ on frequency (see
Eq.(30)).  The numerical values for the absorption cross section
$\sigma_{\nu,H}$ lie in the range $5*10^{-22}$ to $10^{-27}{\rm
cm^{2}}$, with the higher values corresponding to higher frequencies.

It follows that for sufficiently high values of the column density the
optical depth at high frequencies is significantly larger than unity.
Hence the limiting case of high optical depth describes the spectrum,
and the numerical value of the optical depth itself drops out.
Correspondingly we have the case of optical depth much smaller than
unity at low frequencies for all values of column densities
considered.  In this case the actual value of the optical depth does
matter, since the emission is directly proportional to it.
Specifically, the dependence of the absorption cross section on
frequency can be approximated in the FIR/mm-range by a power law.  For
extremely
low values of the column density the optical
depth becomes smaller than unity at all frequencies, and the local
maxima and minima of the absorption cross section then determine the
shape of the emission spectrum directly and produce the various maxima
and minima that we described.  Should some observed spectra show such
features, then we could derive estimates for the column density.
\par
An upper limit for such an estimate can be obtained from the
observationally
inferred gas masses for the molecular gas associated with the dust.
The relationship is:

$${M_{Gas}\over M_{\odot}} = {1\over M_{\odot}}\,\int_{r_0}^R 1.4\,
N_H\,m_p\,2\pi\,r\;dr.\eqno\autnum$$
Observations suggest that the
molecular gas mass is of order $10^{10}\,M_{\odot}$ (see Chini et al.
(1989a)).  This leads to an estimate of the column density of:

$$N_{H,0} \simeq 10^{24}{\rm cm^{-2}},
\eqno\autnum$$
depending on the scale radius $r_c$.\par

\titled{b) Variation of the radial dependence of the column density}\par

\noindent The column density also changes by varying  the radial dependence.
Another possibility is to assume the column density a general powerlaw
dependence on the sperical radius $r$:

$$N_H(r) = N_{H,0}\,(r/r_c)^{-\delta}. \eqno\autnum$$

This variation likewise influence the optical depth. Because the emission
is directly proportional to the actual value of the optical depth,
there are only variation in the IR--range, similar to the case discussed
above if the scale of the column density is small.

\titlec{Variation of the amount of energy transformed into dust heating}

In this paper we simply assume that a fixed fraction of the energy
of the impinging particles is deposited as heat.
Changing this constant factor gives only a variation in the amount of the
calculated IR--luminosity. But in this implies another particles physic
and this is beyond the outlet of the paper.

\titlec {Variation of the dust model}

Recent paper (Sellgren 1981, 1984, L$\acute {\rm e}$ger \& Puget 1984)
show that there is strong evidence
for very small grains and Polycyclic Aromatic Hydrocarbon (PAHs)
in the interstellar medium.
D$\acute {\rm e}$sert et al. (1990) use an empirical dust model made of three
components: big grains, very small grains and PAHs for interprating the
interstellar extinction and the IR emission. They found that the PAH molecules
do not produce any noticeable
absorption features in the NIR extinction curve, although they dominated
the NIR emission (Puget \& L$\acute {\rm e}$ger 1989).
As discussed in the subsection (3.2.2) we are currently independent of
particular
dust grain size distribution.
Beside this we use high values of the column density the optical depth at high
infrared frequencies is significantly greater unity. Hence the limiting case
of high optical depth describes the spectrum in the IR--range,
and the numerical values of the optical depth drops out.
Because of this the exactly from of the extinction curve, especially is
irrelevant. We do not give any prediction for observable spectral features,
which are not generally observed in quasars.

\titleb {Fit to observations}

In this subsection we will now use the model spectra to compare with
the observations.  Here we use the FIR/IR data listed in Neugebauer et
al. (1986), some 3.7$\mu m$ data from Sanders et al. (1989a)
and the mm/submm-data given in Chini et al. (1989a,b).
\par
For sake of consistency we use the same parameters for the
cosmological parameters as Chini et al. (1989a) use:

$$H_0=75{\rm
{km \over s}\,{1 \over {Mpc}}}\;\;\;\;\;q_0=0.5 \eqno\autnum$$
It is pointed out,
that of the nine possible combinations of point or line source
with different z-dependencies for the intensities, and the three
different radial dependencies for the diffusion coefficient only two
combinations give model spectra in the IR compatible with the observed
range:

A line source with an intensity $z^{-1}$ with $D
\sim r \, {\rm or} \sim r^2$.
The case of a constant diffusion coefficient can be ignored.

Please note, that the measured points at $1.3$ mm are usually upper
limits. In the figure captions the parameters used are indicated.

In Fig.(11) the observational data for the quasars 1613+658
and PG1543+658 are fitted with a model spectra.
A line source has been used with an intensity $z^{-1}$
and a minimal distance $z_{\rm min}$ from the center of symmetry
of 0.1pc.

\begfig{5.5}cm{\figure{11}{Observational data from PG1543+489 and
1613+658 with a
line source model spectra with an intensity $^{-1}$ and $D(r)\sim r$.
Parameters: $L_{z,1543}=1.875*10^{47}\,{\rm erg/s}$,
$L_{z,1613}= 5*10^{46}\,{\rm erg/s}$, $N_{H,0} = 10^{24}\,{\rm cm}^{-2}$,
$z_{\rm min} = 0.1\,{\rm pc}$.}}\endfig

The diffusion coefficient is proportional $r$.
The only difference in the model spectra is the source luminosity.
So the model spectra are only shifted in the $L_{\nu}-{\nu}$--diagram
to fit the observations.

\begfig{5.5}cm{\figure{12}{Observational data from 1440+356
and PG1126-041
with a line source model spectra with an intensity $z^{-1}$
and $D(r) \sim r$. Parameter: $L_{z,1440}=1.75*10^{46}\,{\rm
erg/s}$, $L_{z,1126}= 1.25*10^{46}\,{\rm erg/s}$,
$z_{\rm min}=0.1\,{\rm pc}$, $N_{H,0}=10^{24}\,{\rm cm}^{-2}$}}
\endfig

\begfig{5.5}cm{\figure{13}{Observational data from 1449+588
and PG0906+48
with model spectra of line sources with an intensity $z^{-1}$ and
$D(r)\sim r$.
Parameters: $L_{z,1449} = 4*10^{46}\,{\rm erg/s}$,
$L_{z,0906}=1.25*10^{46}\,{\rm erg/s}$, $N_{H,0}=10^{24}\,{\rm cm^{-2}}$
, $z_{\rm min}=0.1\,{\rm pc}$}}
\endfig

In Fig.(12) and (13) the quasars 1440+356, PG1126-041,
1449+588 and PG0906+48 are fitted by models.
Again we use a line source with an intensity $z^{-1}$ and a
diffusion coefficient proportional $r$.
The observations are also fitted by varying the source luminosity.

Note that the observations at 1.3mm is an upper limit. The model
could be shifted closer to the upper limit
at 1.3mm with an adjustment of the scale of the column density
$N_{H,0}$ or the turnover can shifted to higher frequencies,
as long as the scale of the column density is larger then a critical
value ($N_{H,0} \ge 10^{23}{\rm cm}^{-2}$; see the discussion in the
section above).

In Fig.(14) another possible fit is seen, where a smaller scale of
the column density is used.
The quasars PG0050+124 and PG1700+518 are also fitted by a line source
with an intensity $z^{-1}$ and a diffusion coefficient $D\sim r$.

We conclude that all observed spectra can be acceptably fitted with
our model. If it is possible to find another range of parameters which
can fit the observations, the combinations are pointed out
in the subsection above.
For example, if we decrease the minimum distance $z_{\rm min}$
we must increase the source luminosity $L_z$ by a scale factor
to reproduce the same spectrum.

\begfig{5.5}cm{\figure{14}{Observational data from PG
0050+124 and
PG1700+518 with model spectra of a line source with an intensity
$z^{-1}$ and $D(r) \sim r$.
Parameters: $L_{z,0050}=4.375*10^{46}\,{\rm erg/s}$,
$L_{z,1700}=1.3*10^{47}\,{\rm erg/s}$, $z_{\rm min}=0.1\,{\rm pc}$,
$N_{H,0}=5*10^{23}\,{\rm cm^{-2}}$}}\endfig

\begfig{5.5}cm{\figure{15}{Observational data from Mkn231 with a
model spectra of a line source with an intensity $z^{-0.5}$ and
$D(r) \sim r$.
Papameters: $L_{z,Mkn231}=1.2*10^47\,{\rm erg/s}$, $z_{\rm min}
=0.1\,{\rm pc}$, $N_{H,0}= 10^{24}\,{\rm erg/s}$}}\endfig

In Fig. (15) we have tried to fit Mkn231. We use the mm/submm--data
given in Kr\"ugel et al. (1988). In contrast to the other observations
show this radioweak quasar a much steeper decline in the IR--range,
analogous to most of all Markarian galaxies observed in the paper.

For modelling this form of spectra we must use a line source with
an intensity $z^{-0.5}$. We can not clearly restrict the radial behaviour
of the diffusion coefficient, see the discussion
in the subsection (3.3) above, we employ $D(r)\sim r$.

Although the data are well fitted, another possibility exists
to explain the spectrum of Mkn 231. Since most of the Markarian galaxies
of the sample in Kr\"ugel et al. (1988) show a steep decline in the
IR--range and half of them are starburst galaxies, the possibility arises
that in the case of Mkn231 the heating by the starburst dominates.
Any starburst is flat or of positive slope between 100 $\mu$m and 60$\mu$m,
and steeper between 60$\mu$m and 25$\mu$m, as compared to radioweak
quasars.
Besides the ratio of the FIR to radio emission of Mkn 231 is similar to that
for
starburst galaxies (S(60$\,\mu$m)/S(5GHz)$\simeq 200$).
\par

\titlea {Interpretation}

The spectra we selected from the observational database, were chosen
with the following criteria in mind:

\item{--} The entire range of observed luminosities ought to be
represented.

\noindent Hence we can generalize from our fits to these selected
spectra:\par

\noindent For the entire range of observed spectra can we find good
model fits.
Only a small subset of the model parameter range produces
acceptable fits, thus strongly constraining the models.
\item{--} All observed spectra with a straight IR spectrum can be
fitted with a model spectrum using a line source with an intensity $z^{-1}$
with a diffusion
coefficient which is proportional to $r$ or $r^2$.
It should be pointed out that line sources with $\alpha_1 \ge 1$
can be ignored because the decline in the IR--range is too flat.
Additional combinations with $\alpha_1 \le 1$ are possible in some cases,
but not necessary in this context.
\item{--} Apart from the bend near 5*10$^{12}$Hz, which is seen in
some spectra, most observed spectra are quite smooth in the IR.
In our treatment of the IR-radiative transfer we ignore structure
perpendicular to the disk and this
implies a minimal scale of the hydrogen column density:

$N_{H,0} \ge 10^{23}\,{\rm cm}^{-2}.$
\item{--} The observed IR--luminosity is either a factor of 2--3 (for plasma
waves) or 6--10 (for purely hadronic interaction)
smaller than the source luminosity. So the source luminosity can be limited
from the observations, dependent on initial interactions (hadronic cascades
or plasma waves):

$5*10^{47}{\rm erg/s} \le L_Q \le 10^{45}{\rm erg/s}.$
\item{--} The location of the source on the z--axis is irrelevant,
as long as the distance is smaller than a maximal distance:

$z_{\rm min} \le 0.1{\rm pc}.$

\noindent Our conclusions can be summarized by noting that we have a
basic model which appears capable of fitting most, if not all observed
spectra:

Nearly all observed spectra can be fitted with a line source with an intensity
$z^{-1}$.
We can not give a clear restriction to one particular
radial behaviour of the diffusion coefficient, because there is only a small
difference in the decline in the IR--range of the model spectra.
We have either a diffusion coefficient
with a radial dependence $D \sim r^2 \, {\rm or} \sim r$.
Therefore the basic model is a line source, with an intensity
$z^{-1}$ at some indetermined
distance above the the center of symmetry - as long as this distance
is not larger than of order $0.1$ pc - and a diffusion coefficient
which is $D \sim r$ or $\sim r^2$.

It is fairly obvious, that in this model the detailed geometry of the
disk is not important, warp or no warp. The diffused particle distribution has
an
inherent length scale of order the radius of the system and thus a warp of the
sink for the particles does not greatly change the radial dependence of the
particle distribution, unless the disk, i.e. the sink layer, is wrapped around
the primary source.  Thus, warps are certainly not required by our model, but
even if they were to exist, like in Cen A, they would not influence the result
noticeably.
\par
The source energy goes about equally into the central region, where the disk is
too hot for dust, and the outer region, where infrared emission is caused.

A line source with source strength proportional to $z^{-1}$ can be understood
as follows: The energetic particle population has an energy density which
decreases along the jet in the most simple-minded model (see, e.g. Blandford
and K\"onigl 1979) with

$$\epsilon_{\rm rel} = {\rm const.} z^{-2}.\eqno\autnum$$
The lateral losses to the galactic halo per length along the jet then are
proportional to the circumference $2 \pi z \phi$, where $\phi$ is the opening
angle of the jet, and also proportional to the transport through the surface
of the jet:

$$\sim \;\kappa \, {{\partial \epsilon_{rel}} \over {\partial \phi}}
.\eqno\autnum$$
{}From similarity arguments (see subsection 3.1) we infer that the lateral
gradient
scale
and also the transport coefficient are proportional to $\phi z$, and so their
dependencies cancel.  The remaining terms are thus the energy content
multiplied by the circumference and a constant and hence the source strength
is proportional to

$$\epsilon_{rel} \, 2 \pi z \; \sim \; z^{-1}.\eqno\autnum$$
We are thus consistent.
As proven in the appendix the exact numerical value of the diffusion
coefficient drops out of the relations, i.e. is not relevant for the result.
\par

\titlea {Limits of the model}

Here we want to discuss the limitations of the model as presented
above.

The first limitation derives from the assumption that a fixed fraction
of the particle energy flux impinging on the disk is actually
transformed into thermal heating.  In the outer parts of the disk this
assumption may well not be satisfied, since the optical depth for
p-p-interaction - even allowing for the random walk of charged
energetic particles - is too low.  On the other hand, since this is
likely to happen in that radial range where heating from other
physical causes like Supernova induced energetic particles takes over
this is unlikely to change the resulting FIR/submm/mm-spectrum.

A second limitation derives from the fact that we have not specified
the exact physical mechanism to transform the impinging energetic
particle flux into heat.  Assuming that this is solely via
p-p-interactions leads to approximately equal luminosities in the
three channels FIR, Gamma-rays and neutrinos.

However, as we pointed
in the beginning of this paper, strong p-p-interactions in a dense
molecular cloud lead to a strong particle density gradient form
outside to inside the cloud and thus to strong wave excitation
(Skilling \& Strong 1976); these waves in turn strongly dissipate inside the
cloud in ion-neutral friction processes and thus heat the cloud (Kulsrud \&
Pearce 1969).  In
this case, that part of the disk which consists of many dense
molecular clouds with lots of dust, may get heated by this second
process, while the inner most part of the disk, which is homogeneously
warm and hot, gets directly heated by pp-interactions and the ensuing
cascades.

The diffuse extragalactic radiation background in the FIR and
Gamma-range (see Ressel \& Turner 1990) clearly demonstrates that there the
Gamma-ray range is much weaker than the X-ray and FIR range.  Hence, taking a
clue from recent ROSAT-results, which suggest that at least in X-rays the
background is mostly made up by extragalactic point sources, viz. AGN, then the
overall spectrum of even a single AGN should not typically have equal
luminosities in the three wavelength bands FIR, X-ray and Gammas.  With
the composite heating mechanism discussed above this will indeed not
be the case:

The heating to produce the FIR is independent of which of the two
simple mechanisms is valid, while the X-rays arise only as a
consequence of the p-p-cascades and interaction in the disk.
We will discuss the consequences of this picture elsewhere, both for
quasars and active galactic nuclei in general as well as for the inner
region of our Galaxy.

Another way to change the relative weight of the three channels FIR,
X-rays and Gammas in the p-p-interaction mechanism, is a variation of
the particle spectrum.  For a spectrum steeper than $E^{-2}$ we reduce
the Gammas and increase, relatively, the heating from simple pair
production (lower threshold).

We have ignored possible loss processes, like p-Gamma and adiabatic
losses during the diffusion of the energetic particles from the
central source out through the tenuous medium to the disk.  This will
be an interesting question to consider elsewhere.\par

\titlea {Conclusions and extrapolations}

In this paper we have developed a model that explains the FIR spectra
of radioweak quasars as resulting from heating by energetic particles
which are produced in a putative source, such as a jet, on the symmetry axis at
a small, but basically undetermined distance from the center of symmetry, the
central engine (only condition: distance less than about a 0.1 parsec).

We find that for reasonable parameters we can indeed reproduce
the observed spectra.  We find that for a diffusion coefficient in the
region above the disk, which scales with radius $r$, and a line source strength
$
\sim \, z^{-1}$, with a total source luminosity of $2$ to $3$ (for plasma
waves
interaction) or $6$ to $10$ (for purely hadronic initial interaction) times the
observed infrared luminosity, we can reproduce and interpret the spectra of
radioweak quasars from the mm to the near infrared region.

There are consequences of our model for the
X-ray, Gamma- and neutrino emission from the inner disk, as well as
likely consequences for the heating of interstellar clouds close to a
central engine (by wave excitation and dissipation), which remain to be worked
out.
The analogy to the Galactic center also suggests that the accretion
luminosity scales with the jet luminosity, and with the FIR emission,
but may be larger by a factor between 1 to 10. Such correlations will provide
a severe test for the proposed speculative model.
\par

\ack {We wish to thank J.S. Mathis, E. Kr\"ugel and R. Chini for discussion
of radiative tranfer in the IR.  We wish to acknowledge also discussions with
W.
Kr\"ulls, R. Schaaf, K. Mannheim, S. Linden, H. Falcke and J. Rachen.  High
energy physics with PLB is supported by DFG Bi 191/6/7/9 (Deutsche
Forschungsgemeinschaft), the BMFT (FKZ 50 OR 9202) (Bundesministerium f\"ur
Forschung und Technologie, through the DARA), and a NATO travel grant.}

\begref

\ref Alloin D., Barvainis R., Gordon M.A., Antonucci R.R.J., 1992,
     A\&A 265, 429
\ref Barvainis R., 1990, ApJ 353, 419
\ref Barvainis R., Alloin D., Antonucci R.R.J., 1989, ApJ 337, L69
\ref Biermann P.L., Harwit M., 1980, ApJ 241, L105
\ref Biermann P.L.,  1992 in "High energy neutrino astrophysics", Eds. V.J.
     Stenger J.G., Learned S., Pakvasa X., Tata, World Scientific, p. 86
\ref Biermann P.L., Strittmatter P.A.,  1987, ApJ 322, 643
\ref Biermann P.L., 1993, A\&A 271, 649, (paper CR I)
\ref Biermann P.L., Cassinelli J.P., 1993, A\&A (in press,
     paper CR II)
\ref Biermann P.L., Strom R., 1993 (in press, paper CR III)
\ref Blandford R.D., K\"onigl A., 1979, ApJ 232, 34
\ref Camenzind M., 1990, Rev.Mod.Astron. 3, 234
\ref Chandrasekhar S., 1943, Rev.Mod.Phys. 15, 1
\ref Chini R., Kreysa E., Biermann P.L., 1989a, A\&A 219, 87
\ref Chini R., Biermann P.L., Kreysa E., Gem\"und, H.-P., 1989b,
     A\&A 221, L3
\ref Chini R., Kr\"ugel E., Kreysa E., 1986, A\&A 167, 315
\ref Clavel J., Nandra K., Makino F., Pounds K.A.,
     Reichert G.A., Urry C.M., Wamstecker W.,
     Peracaula-Bosch M., Stewart G.C., Otani C., 1992, ApJ 393, 113
\ref de Kool M., Begelman M.C., 1989, Nat 338, 484
\ref D$\acute {\rm e}$sert F.-X., Boulanger F., Puget J.L., 1990, A\&A
     237, 215
\ref Dickel J.R., Breugel W.J.M. van, Strom R.G., 1991, AJ 101, 2151
\ref Drury L.O'C., 1983, Reports on Progress in Physics 46, 973
\ref Edelson R.A., 1986, ApJ 309, L69
\ref Falcke H., Mannheim K., Biermann P.L., 1992, A\&A (submitted)
\ref George I.M., Fabian A.C., 1991, MNRAS 249, 352
\ref Hildebrandt S.E., Whitcomb S.E., Winston R., Stiening R.F., Harper
     D.A., Moseley S.H., 1977, ApJ 216, 698
\ref Kamke E., {\it Differentialgleichungen}, ed. B.G. Teubner, 1961
\ref Kronberg P.P., Biermann P.L., Schwab F.R.,  1981, ApJ 246, 751
\ref Kronberg P.P., Biermann P.L., Schwab F.R.,  1985, ApJ 291, 693
\ref Kr\"ugel E., Chini R., Kreysa E., Sherwood W.A., 1988, A\&A 190, 47
\ref Kulsrud R.M., Pearce W.P., 1969, ApJ 156, 445
\ref Lacy M., Rawlings S., Hill G.J.,  1992, MNRAS 258,
     828
\ref Lawrence A., Rowan-Robinson M., Efstathiou A., Ward M.J.,
     Elvis M., Smith M.G., Duncan W.D., Robson E.I., 1991, MNRAS 248, 91
\ref L$\acute {\rm e}$ger A., Puget J.L., 1984 A\&A 137, L5
\ref MacDonald J., Stanev T., Biermann P.L., 1991, ApJ 378, 30
\ref Mathis J.S., Rumpl W., Nordsieck K.H., 1977, ApJ 217, 425
\ref Mezger P.G., Mathis J.S., Panagia N., 1982, A\&A 105, 372
\ref Miley G.K., Neugebauer G., Clegg P.E., Harris S.,
     Rowan--Robinson M., Soifer B.T., Young E., 1984, ApJ 278, L79
\ref Miley G.K., Neugebauer G., Soifer B.T., 1985, ApJ 293, L11
\ref Nellen L., Mannheim K., Biermann P.L.,  1993, Phys.Rev. D (in press)
\ref Neugebauer G., Becklin E.E., Oke J.B., Searle L., 1979,
     ApJ 230, 79
\ref Neugebauer G., Miley G.K., Soifer B.T., Clegg P.E., 1986,
     ApJ 308, 815
\ref Niemeyer M., 1991, M.Sc. Thesis, University of Bonn
\ref Parker E.N., 1958, ApJ 128, 664
\ref Puget J.L., L$\acute {\rm e}$ger A., 1989, Ann.Rev. A\&A 27, 161
\ref Rawlings S., Saunders R., 1991, Nature 349, 138
\ref Ressel M.T., Turner M.S., 1990, Comments Astrophys. 14, 323
\ref Rieke G.H., Low F.J., 1975, ApJ 199, L13
\ref Roche P.F., Aitken D.K., Phillips M.M., Whitmore W., 1984,
     MNRAS 207, 35
\ref Sanders D.B., Scoville N.Z., Soifer B.T., 1988, ApJ 335, L1
\ref Sanders D.B., Phinney E.S., Neugebauer G., Soifer B.T., Matthews K.,
     1989a, ApJ 347, 29
\ref Sanders D.B., Scoville N.Z., Zensus A., Soifer B.T., Wilson T.L.,
     Zylka R., Steppe H., 1989b, A\&A 213, L5
\ref Schlickeiser R., Biermann P.L., Crusius-W\"atzel A., 1991, A\&A 247,283
\ref Sellgren K., 1981 ApJ 245, 138
\ref Sellgren K., 1984 ApJ 277, 623
\ref Skilling J., Strong A.W., 1976, A\&A 53, 253
\ref Spitzer L.Jr., Tomasko M.G., 1968, ApJ 152, 971
\ref Stein W.A., Soifer B.T., 1983, Ann. Rev. A\&A  21, 177
\ref Telesco C.M., Harper D.A., 1980, ApJ 235, 392
\ref Wang Y.M., Schlickeiser R., 1987, ApJ 313, 200
\ref Wilson A.S., Elvis M., Lawrence A., Bland-Hawthorne J.,  1992 ApJ 391,
     L75
\endref
\vskip 2cm
\app{A}
\titled{A.1 The solution of the partial diffusion equation}
For calculating the energy deposition rate as a function of the
particles energy (Eq.(5)), we must solve the diffusion equation which
is a partial differential equation:
$$\eqalignno{-Q/C1 &= \beta
r^{\beta-1}\cr &+ r^{\beta-2}{\partial \over \partial
r}\left(r^2{\partial W\over \partial r}\right) +{\partial \over
\partial \mu} (1-\mu^2){\partial W \over \partial \mu} &\autnumA
\cr}.$$
We use the scattering time method (e.g., Wang, Schlickeiser,
1987):

$$W(r,\mu) = \int _0^\infty f(r,a)g(\mu,a) \;da \eqno\autnumA$$
This method can always be used when the differential operator can be
separated in variables.

Inserting the above expression into the partial differential equation
yields:

$$ \int_0^\infty g\left(\beta
\,r{\partial f\over \partial r}+{\partial \over \partial r}(r^2{\partial f
\over \partial r})\right)$$

$$+f\left({\partial \over \partial \mu} (1-\mu^2){\partial g\over
\partial \mu}\right)\;da =-{Q\over D_1}r^{2-\beta}.
\eqno\autnumA$$

The functions $f(r,a)$ and $g(\mu,a)$ have to obey the following
conditions:

$${\beta \,r\,{\partial f\over \partial r}+{\partial \over \partial r}
(r^2\,{\partial f\over \partial r})} = {\partial f\over\partial a},
\eqno\autnumA$$

$${\partial \over \partial \mu }(1-\mu^2){\partial g\over \partial
\mu} = {\partial g\over\partial a}.\eqno\autnumA$$
Putting this into the the basic differential equation again and
integrating yields:

$$f(r,+\infty)g(\mu,+\infty)-f(r,0)g(\mu,0)$$ $$ =-{1\over 2\pi z_0^2
C_1} r^{2-\beta}
\delta(\mu-1)\delta(r-z_0).\eqno\autnumA$$
This is equivalent to the following boundary conditions:

$$\eqalignno{f(r,+\infty)g(\mu,+\infty) &= 0,\cr f(r,0)& = {1\over
2\pi z_0^2 D_1}r^{2-\beta}\delta(r-z_0)\cr g(\mu,0) &=\delta(\mu -1).&
\autnumA \cr}$$
This means that the new equations with the new boundary conditions are
equivalent to the earlier partial differential diffusion equation with
its boundary condition.  We use the following ansatz:

$$ g(\mu ,a) = \sum_{l =0}^\infty
\exp(-\lambda_l^2\,a)\,h_l(\mu).\eqno\autnumA$$
It follows, that:

$${\partial \over \partial \mu}\,(1-\mu^2)\,{\partial h_l\over
\partial
\mu}+\lambda_l^2\,h_l(\mu) =0.\eqno\autnumA$$
This is the Legendre differential equation, where the
eigenva\-lues $\lambda_l$ are given by $\lambda_l = l(l+1)$ mit $l\in
{\bf N}$. Solutions are obviously the Legendre polynomials:

$$h_l(\mu ) = C_l\,P_l(\mu).\eqno\autnumA$$
The full solution is then:

$$g(\mu ,a) = \sum_{l=0}^\infty C_l\,\exp(-l(l+1))\,P_l(\mu).
\eqno\autnumA$$
Since we have a fully absorbent boundary on the disk, the eigenvalues
$l$ are restricted.  It follows that we have only uneven terms $l$,
i.e.$l=2n+1$.  The constant $C_l$ also follows from boundary
conditions.

$$\Longrightarrow C_l\,P_l(\mu ) = \delta(\mu -1). \eqno\autnumA$$
For the solution we use the orthogonality of the Legendre polynomials:

$$\int_0^1 P_m(x)\,P_n(x)\;dx = {1\over 2m+1}\,\delta_{mn}
\eqno\autnumA$$

$$\Longrightarrow {1\over 2l+1}\,C_l = \lim_{\epsilon \to 0} \int_0^1
\delta(\mu-(1-\epsilon))P_l(\mu)d\mu
\eqno\autnumA$$
Integrating and going to the limit of $\epsilon \to 0$ we obtain for
$C_l$:

$$C_l = 2l+1 =4n+3.\eqno\autnumA$$
As the next step we we use the
ansatz for $g(\mu ,a)$:

$$W(r,\mu ) = \sum_{n=0}^\infty
h_{2n+1}(\mu)\,t_{2n+1}(r),\eqno\autnumA$$
with

$$t_{2n+1}(r) = \int_0^\infty f(r,a)\,\exp(-(2n+1)(2n+2)).
\eqno\autnumA$$
After multiplication of our result for the function $g$ with the
equation for $f$ and integration we obtain:

$$\int_0^\infty {\partial
f\over \partial a}
\sum_{n=0}^\infty h_{2n+1}(\mu)\,\exp(-(2n+1)(2n+2))\;da$$

$$=O_r(f(r,a)) \,\sum_{n=0}^\infty
h_{2n+1}(\mu)\,\exp(-(2n+1)(2n+2))\;da, \eqno\autnumA$$
where $O_r$ is
the $r$-dependent differential operator.  Applying (A17) and partially
integrating gives:

$$\sum_{n=0}^\infty h_{2n+1}(\mu)\left( -f(r,0)\right)+(2n+1)(2n+2)\,
h_{2n+1}(\mu)\,t_{2n+1}(r)$$

$$=\sum_{n=0}^\infty
h_{2n+1}(\mu)\,O_r(t_{2n+1}(r)) \eqno\autnumA$$

$$\Longleftrightarrow r^2\,{\partial^2 t_{2n+1}\over \partial r^2}+
(2+\beta )\,{\partial t_{2n+1}\over \partial
r}-(2n+1)(2n+2)\,t_{2n+1}$$

$$ = -{\delta(r-z_0)r^{2-\beta}\over 2\pi z_0^2 D_1}\,Q_2(E), \eqno\autnumA$$
Hence we have for the radial dependence of the functions $t$ an
ordinary differential equation. The homogeneous solution is (Kamke
1961):

$$ t_{2n+1}^1(r) = r^{1/2-a/2+d/2}\eqno\autnumA$$

$$ t_{2n+1}^2(r) =
r^{1/2-a/2-d/2}\eqno\autnumA$$
with
$$a=2+\beta\,\,\,\,b=(2n+1)(2n+2),\eqno\autnumA$$

$$d=\sqrt{(1+\beta)^2+4(2n+1)(2n+2)}.\eqno\autnumA$$
It follows that:

$$t_{2n+1}^1(r) = r^{-1/2-\beta/2+d/2}= r^{d_1},\eqno\autnumA$$

$$t_{2n+1}^2(r) = r^{-1/2-\beta/2-d/2}= r^{d_2}.\eqno\autnumA$$
The
solution of the inhomogeneous equation can now be derived with the
source function of the adjoint equation:

$$ t_{2n+1}(r) =
\int_0^\infty G(r,r_*)\,Q_{adj}(r_*)\;dr_* ,\eqno\autnumA$$
with the
Green-function

$$G(r,r_*) = {1\over J}
\cases{t_{2n+1}^2(r_*)\,t_{2n+1}^1(r),&if $0\le r\le r_*$\cr \cr
t_{2n+1}^1(r_*)\,t_{2n+1}^2(r),&if $r_*\le r\le
\infty$\cr}
\eqno\autnumA $$

$J$ is composed from the Wronski-determinant {\bf W} and the first
coefficient of the differential operator in Sturm-Liouville-form {\bf
$\Psi$}:

$$\eqalignno{ J &= {\bf W}(t_{2n+1}^1(r),t_{2n+1}^2(r)){\bf
\Psi}
\cr & \cr &= -\sqrt{(1+\beta
)^2+4(2n+1)(2n+2)}.&\autnumA \cr} $$
After integration we find the
solution of the inhomogeneous ordinary differential equation:

$$t_{2n+1}(r) =-{1\over 2\pi d D_1}\,Q_2(E) \cases{
z_0^{d_2}\;r^{d_1}, &if $0\le r\le z_0$\cr \cr z_0^{d_1}\;r^{d_2}, &if
$z_0\le r\le \infty.$\cr}
\eqno\autnumA$$
This leads to a full solution of our basic diffusion equation:

$$W(r,\mu ) = -\sum_{n=0}^\infty {C_{2n+1}P_{2n+1}(\mu) t_{2n+1}(r)}
\eqno\autnumA$$
This is the distribution function, from which we determine the
deposition rate of the particles in the disk.
\par
When we take the derivative of this distribution and multiply with the
diffusion coefficient, to calculate the rate at which energy is deposited in
the
disk, the value of the diffusion coefficient drops out.
 \bye